\documentclass[12pt]{article}
\usepackage[dvips]{graphicx}
\title{
\hspace{10 cm}{\small ISU-IAP.TH 99-02 IRKUTSK.}\\
\vspace{1 cm}
 THE PRODUCTION OF LIGHT GOLDSTONE PARTICLES ON PHOTON COLLIDERS\\
}
\author{S.I.Polityko \\
\it {IRKUTSK STATE UNIVERSITY}\\
E-mail polityko@psi.isu.ru
}

\begin{document}
\language=1
\maketitle
\begin{abstract}
        It is shown that by realizing the project of intensive
$\gamma $ beams with large energy (project PLC) an essential flux
of light Goldstone particles (axions, arions, familons, majorons)
will be generated. The light higgs can be observed via interaction
with matter. The probability of light higgs - electron production
by absorption of several laser photons simultaneously is calculated.
\end{abstract}

\section {Introduction}
        The most important peculiarity of electroweak interaction theory
is its renormalizability which is provided by Higgs boson introduction.
Their search is considered to be one of the most important problems of
elementary particle physics for the nearest future.
In fact, there is no special reason to consider that the Higgs sector
contains the only one boson. There are some theoretical considerations
pointing out the desirability of Higgs sector expanding \cite{Ans}.
In this case in different variants of the theory together with the heavy
particles there appear the light pseudogoldstone bosons.
In the paper we propose the investigation of new possibilities to
study nonstandard Higgs particles such as axion, arion, majoron,
familon on the basis of colliding photon-electron bunches.

        Nowaday the experiments on the colliding beams have become
the source of fundamental information about the matter.
On the basis of the future linear colliders (Next Linear Collider)
one can make  the study of reaction region wider.
Lasers available nowadays permit to get the intensive dense photon
beams at Compton scattering on the electronic beams
\cite{GKST1,GKST2,GKPST,Gin,Tel1,Tel2,Brink}.
The obtained $\gamma e$ and $\gamma \gamma $ bunches have the same energies
and luminosities as the basic electron bunches.
For obtaining the intensive photon beams one can use the solid state laser
and also the free electron laser.
It allows one  to study experimentally new problems inaccessible for
investigation by other methods.

First, the conversion region is the region of extremely intensive
electromagnetic field (a focused laser bunch).

Second, the conversion region can be treated as the
$\gamma _0 e$ - collider ($\gamma _0$ is the laser photon).

Though the energy of the $\gamma _0 e$ system (in the center-mass
system) is not too large , the luminosity of this collider is very large.
Let $E_e$ be the energy of electron in the beam, $\omega _0$
is the energy of laser photon.
Then $$W_{\gamma _0 e}
= m_e \sqrt {1+x},\quad x=4E_e\omega _0 / m_{e}^2=15.3
\biggl [{E_e\over TeV}\biggr ]\biggl [{\omega_0 \over eV}\biggr ].
$$
Note via  $k$-- the conversion coefficient  $ e \to \gamma $,
$\sigma $-- the cross section of the Compton scattering,
$N_e$ -- the number of electrons in a bunch, $f$--the repetition rate,
 $S$-- the transverse cross section of the electron bunch in the conversion
region, $N_{\gamma _0}$- the number of laser photon going through the
electron bunch.
Then the luminosity is
\par
$$
L_{{\gamma}_0 e} = f{N_e N_{\gamma _0 e} \over S}
$$
\par
\noindent
At the same time,  $k \approx N_{\gamma _0} \sigma /S$.
Therefore
$$
L_{{\gamma}_0 e} = f\cdot {N_e k \over \sigma} = 10^{38} cm^{-2} c^{-1}\eqno(1)
$$
Accordigly, the region of laser conversion gives
us the possibility to investigate the rare processes of creation of
light particles with masses $\leq W_{\gamma _0 e}$ \cite{Pol1,Brod,Pol2,GKP}.

    It is very important to note that these phenomena can be observed
with obtaining high energy photons simultaneously.
They can also be specially studied  on electron beams of existing
accelerators such as LEP, SLC and TRISTAN.

 The region of laser conversion $e \rightarrow \gamma $ is unique in its
physical properties and allows one to study new problems previously
inaccessible for investigation -simply as a by-product of obtaining
$\gamma $ beams.
In particular it is the creation of light pseudo-Goldstone bosons
such as axion, arion, majoron, familon.
Due to high luminosity of $n\gamma_0 e$ system the possible mechanism
of creation is the processes with simultaneous absorption of few $(n)$  laser
photons from laser wave  by an electron:
$$
n\gamma_0 + e \rightarrow X + e \eqno(2)
$$

In the paper we study the mechanism of light Goldstone particles (2)
production\footnote {
Note the second mechanism of production:
$n\gamma_0+\gamma \rightarrow X$, where $\gamma $  is the high energy photon,
but the number of produced light Goldstone particles is essentially
less (contains additional
factor $m_X^2/m_e^2$)}.
This seems interesting from the following points of view:

a)the possibility of experimental observation or the set-up of new
constraints on the couplings of light particles with leptons and photons;

b) the influence of the non-linear effects on the probability of creation.

\section{ The properties of light Goldstone particles
and experimental constraints on their couplings}

\subsection {The "invisible" axion}
      The light pseudogoldstone boson, named axion, was proposed
for the solution of the CP-violation problem in strong interactions
\cite{SW,FW}.
This standard axion is not observed in a number of experiments \cite{GGR}.
The idea of natural CP symmetry is attractive, and the theory of the standard
axion was modified to make it interact more weakly with matter and to
make it lighter.
The axion model with two Higgs doublets is characterized by the scale of
breaking
of the $U(1)$-symmetry $f_{pq} \approx 250$ GeV and the mass $m_a \geq 150$
KeV, connected by one parameter: the ratio of the vacuum expectation value
(VEV) of Higgs doublets.
The introducing of additional Higgs multiplets separates $m_{a}$ and $f_{pq}$,
and $250$ GeV $\leq f_{pq} \leq 10^{19}$, the mass $m_{a}$,as the coupling
with matter can become sufficiently small \cite{Ans}.
 One of such possibilities is the introduction of the additional scalar field
($SU(2)\times U(1)$ singlet) at arbitrary large VEV.
This "invisible" axion is well-known as Dine-Fischler-Srednicki-Zhitnitsky
axion (DFSZ) \cite{Z,DFS}.
The interaction Lagrangian of the axion with the electrons and the photons
has the form
$$
{\cal L}= g_{aee}\bar e \gamma _5 ea+C_{a\gamma \gamma}{\alpha \over m_e}
aF\tilde F \eqno(3)
$$
The couplings $g_{aee}$ and $C_{a\gamma \gamma}$ are given by:
$$
g_{aee}={m_am_e\over f_\pi m_\pi}{1+z\over N\sqrt{z}}{1\over v^2+1},
$$
$$
C_{a\gamma \gamma}={m_am_e\over 8\pi f_\pi m_\pi}{(1+z)\over \sqrt{z}}
({8\over 3}-{2\over 3}{(4+z)\over (1+z)}),
$$
\par
\noindent
where $f_\pi=94$ MeV is the pion decay constant, $z=m_u/m_d=0.568$
is the ratio of the quark masses, $N$ is the number of generations,
$v$ is the ratio of the VEV's of two Higgs fields.
Respectively, the mass of axion is given by
$$
m_a=m_\pi{f_\pi\over f_a}{N\sqrt{z} \over 1+z}(v+1/v).
$$
Another possibility of "invisible" axion introduction was proposed in \cite{K,SVZ},
now this axion is well-known as hadronic or Kim-Shifman-Vainstein-Zakharov
axion (KSVZ).
The KSVZ axion does not interact with leptons at the tree level,
because the couplings of interaction to leptons and photons
are two orders less compared to DFSZ axion.
The archion model can be treated as this class of axion models \cite{Ber}.
This model naturally contains  the global symmetry $U(1)$,
the spontaneous breaking of which leads to the appearance of Goldstone boson ,
which has both diagonal and nondiagonal flavor interaction with fermions.
But unlike the axion the archion has no interaction with
photons and it is like a hadronic axion with strongly suppressed lepton
interaction.

The astrophysical constraints on the mass and VEV of "invisible" axion
 \cite{Kim}:
$$
10^9 GeV < f_a < 10^{12}GeV,\quad \quad 0.6\cdot 10^{-5}eV < m_a <
0.6\cdot 10^{-2}eV
$$
lead to following values of coupling with leptons
($(g_{aee}=m_e/vf_a)$): $g_{aee}=0.5\cdot 10^{-12}-0.5\cdot 10^{-9}.$

    The acceleration data give us more less coupling constraints.
In PDG \cite{PDG} for the Lagrangian
${\cal L}= G_{aee}\partial _\mu a\bar e \gamma_\mu \gamma _5 e$
the corresponding constraint is $G_{aee} < 2.7\cdot 10^{-5} GeV^{-1}$,
that  gives for coupling of Lagrangian (3) ($g_{aee}=2m_eG_{aee}$)
$g_{aee}<3\cdot 10^{-8}.$

\subsection {The Arion}

        The arion is a neutral, strictly massless , stable pseudoscalar
boson with even charge parity, having interactions with fermions \cite{Ans}.
The interaction of an arion with lepton has a form:
$$
{\cal L}=v_1{m_e\over v}(i\bar e \gamma_5 e)\alpha,\quad
v=(G_F\sqrt{2})^{-1/2}=246\quad GeV.
$$
Here  $\alpha$ is the arion field, $v_1$ is the dimensional parameter,
equal to the ratio of different VEVs, which theoretical value is of
the order 1.
However there exist strong restrictions  on this value from astrophysical
considerations.
Weakly interacting with a matter the arions can be emitted from stars.
The arion emission leads to fast loss of energy from the stars.
The demand of the condition by which the arion luminosity of Sun should
not exceed the photon luminosity
leads to $v_1<10^{-3}$.
A more strong constraint appears from the evolution of a red giants:
$v_1<10^{-6}$.
Thus, the coupling with leptons $g_{aee} <2\cdot 10^{-9}- 2\cdot 10^{-6}$.

\subsection { The Œ joron}

    A spontaneously broken global symmetry of lepton number will lead to
massive Majorana neutrinos and a Nambu-Goldstone boson the Majoron. This
Goldstone
can be accomplished by extending the standard model with an additional
$SU(2)-$ triplet Higgs multiplet.
In Gelmini-Roncadelli model \cite{GR}  the triplet of Higgs fields with
small neutral component VEV provide the small majoron mass of a neutrinos.
The Majoron almost consist from neutral triplet component fields,
connected only with a neutrino and consist of small admixture of doublet field.
Because the majoron weakly interact with a leptons.
The Lagrangian of interaction with an electrons has a form
$$
{\cal L}= 2\sqrt{2}G_Fvm_e\bar {e}i\gamma_5 e\Phi_M,
$$
where $\Phi_M $ is the majoron field, $v$ is VEV.
The anomalous coupling of the Majoron to photons vanishes.

The astrophysical constraint from the consideration of Majoron emission
rates from the neutron-star core is $v< 2KeV$.\footnote {In another majoron models the constraint on the coupling
with the leptons more hard $g_{\Phi_M ee}=2\sqrt{2}G_Fm_ev<1.7
\times10^{-18}$.}

Thus, we have the following constraint on the coupling with an electron:
$g_{\Phi ee}<3.4\cdot 10^{-14}$.

\subsection {The Familon}

   The familon is the Goldstone boson associated with the spontaneous
breaking of a global family symmetry (horizontal symmetry) \cite{Cheng}.
The horizontal symmetry is the spontaneous breaking symmetry between
the generation of a quarks and a leptons.
Since the breaking of the horizontal group must take place within small distance
the familon,  similar the invisible axion, interact weakly with the matter
and has small mass.

      Phenomenologically the effective interaction of a familon with a leptons
at low energy can be written in the form:
$$
{\cal L}={1\over F}(2m_e \bar {e}\gamma _5 e)\Phi_F.
$$

 The astrophysical constraint on the coupling:
$F>7\cdot 10^{9}$ GeV,
its correspond $g_{Fee}=2m_e/F<1.4\cdot 10^{-13}$.
The investigation of decay $K^+\rightarrow \pi^+\Phi_F$  gives us
the following constraint
\cite{PDG}
$B(K^+\rightarrow \pi^+\Phi_F) < 1.7\cdot 10^{-9}$ or $F > 1.3\cdot 10^{11}$
GeV.

\section{ The calculations of probability and the number of events.}

     The interaction Lagrangian of the light Goldstone particles with the
electrons in general form can be written as
$$
{\cal L}=g_{Xee}( \bar {e}\gamma _5 e)\Phi_X.
$$
     We can consider the field of laser wave with the circularly polarization.
The vector potential of a electromagnetic wave has the form
 $A_{\mu } = a_{1\mu }\cos (\varphi ) + a_{2\mu } \sin (\varphi ),
 \varphi =kx$ ,where $k_{\mu }$ is the momentum of laser photon,
 ($ka_{1}=ka_{2}=a_{1}a_{2}=0,
 a_{1}^{2}=a_{2}^{2}=a^{2}$).
 The matrix element of the light Goldstone by electron in external field
can be written in the form
    $$
    M_{fi}=ig_{Xee} \int d^{4}x \bar \Psi _{e}(x)
    \gamma _{5} \Psi _{e} \phi _{X}(x), \eqno(4)
    $$
    \par
    \noindent
     where  $\Psi _{e}(x) \quad (\bar \Psi _{e}(x))$ are the exact solution
of the Dirac equation for an electron in the field of a circularly polarized
wave:
    $$
    \Psi _{e}(x)=(1+{ek\hat A \over 2kp})u_{p}\exp
    (ie{a_{1}p \over kp} \sin (\varphi )-ie{a_{2}p \over kp} \cos (\varphi )
    +iqx), \quad \phi _{X}={e^{-ip_{X}x} \over \sqrt{2\epsilon _{X}}}.
    $$
    \par
    \noindent
   Here $\epsilon _{X},p_{X}$ is the energy and the momentum of a Goldstone,
  $p$ and $q$ are the momentum and "quasimomentum"  of an electron
\footnote{The introduction of quasimomentum of an electron take into account
the effective "heavier" of electron in the electromagnetic wave field
$m_{*}^{2}=m_{e}^{2}(1+\xi ^{2})$. It is important for the determination of
production threshold.}:
    $$
    q_{\mu } =p_{\mu }+\xi ^{2}{m_{e}^{2} \over 2kp}k_{\mu },
    \quad \xi ^{2}={e^{2}a^{2} \over m_{e}^{2}}.
    $$
   \par
   Using the standard techniques (see \cite{Ritus,LD}), we calculated
the probability of Goldstone production by nonpolarized electron
$$
dW={g_{Xee}^2 m_e\over 64\pi }\sum W_{n}
                    {du\over (1+u)^2}\eqno(5)
$$
\par
$$
 W_{n}= {1\over \sqrt{nx+1+\xi ^2}}
\left (-2\eta ^2 J_n^2 (z)+\xi ^2 {u^2\over 1+u}[J_{n+1}^2(z)+J_{n-1}^2(z)
                   -2J_n^2(z)] \right ),
$$
\par
\noindent
where $J_{n}(z)$ is the Bessel function of $n$-th order,
$u=(kp_a)/(kp^{'} )$, $p^{'}$ is the momentum of scattering electron,
$$
\eta ={m_X\over m_e},\, z={2\xi \over x}\sqrt{nx+1+\xi ^2}\sqrt{{
(u_+-u)(u-u_-)\over (1+u_+)(1+u_-)}},\,u_-<u<u_+,
$$
\par
$$
u_{\pm}={nx+\eta ^2 \pm \sqrt{(\eta ^2-nx)^2-4\eta ^2(1+\xi ^2)}\over
2+nx+2\xi ^2 -\eta ^2 \mp  \sqrt{(\eta ^2-nx)^2-4\eta ^2(1+\xi ^2)}}.
$$
\par

In formula (5) the term with concrete $n$ corresponds to the Goldstone
production by an electron via absorption from electromagnetic wave $n$
laser photons simultaneously (2),
$n_{th}$ is the minimal number of photons for the Goldstone production
with the mass $m_X$
$$
n_{th}={1 \over x}(\eta ^2 +2\eta \sqrt{1+\xi ^2})\eqno(6)
$$

   Integrating the expression (5) we get the total production probability of
a light Goldstone:
$$
{g_{Xee}^2 m_e\over 64\pi }f(x,\xi,\eta).
$$
In Figure 1 the dependence of function $f(x,\xi,\eta)$ shown at some values
$\xi$.
\begin{figure}
\centering\includegraphics[bb=170 160 430 650,scale=0.5]{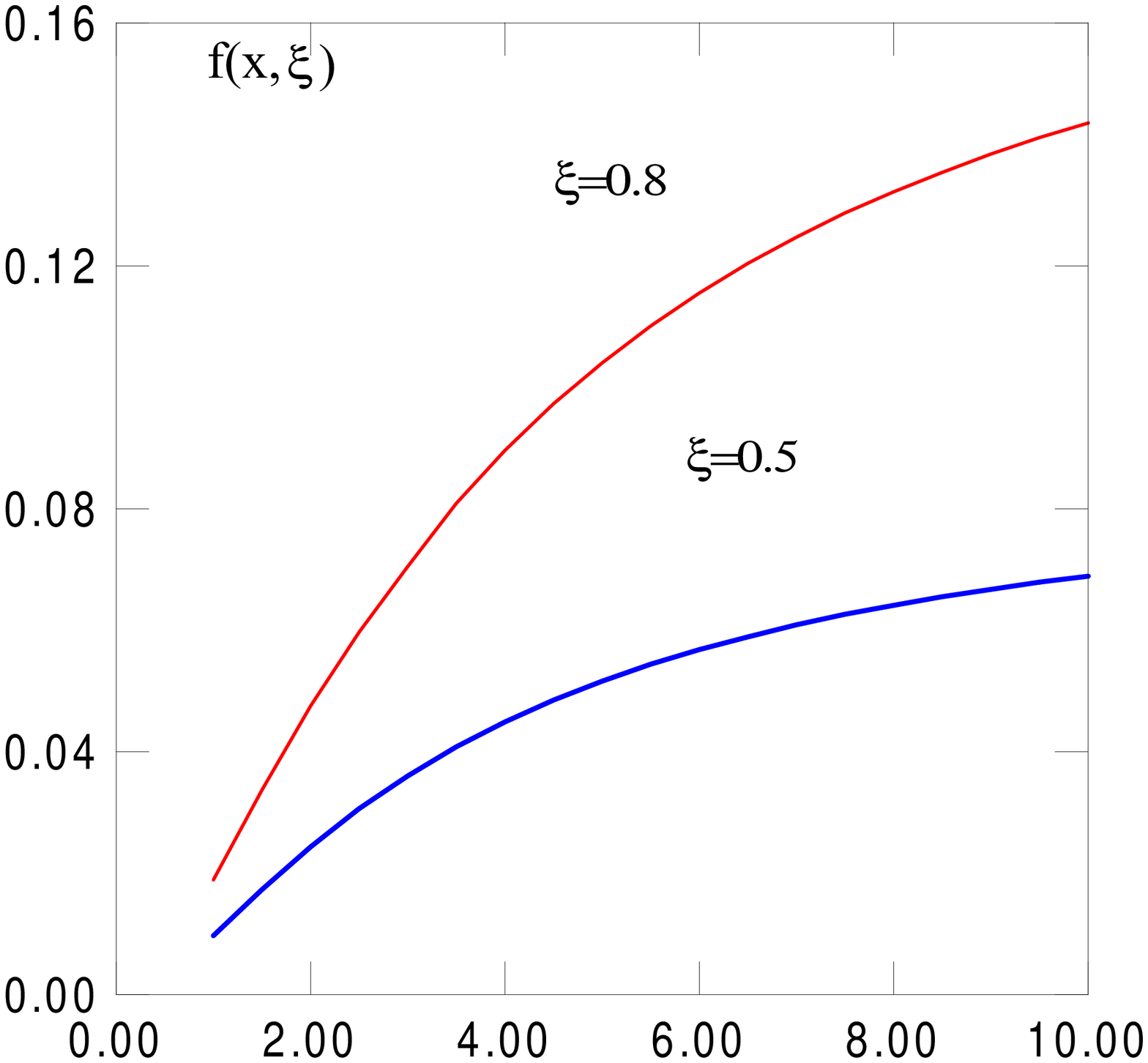}
\caption{The dependence $f(x,\xi)$ from $x$ at some values $\xi$
 and $\eta =0$}
\end{figure}
At $ \xi ^2 \ll 1$, expanding the Bessel function in series, we get from (5)
$$
dW_n={g_{Xee}^2m_e(\xi /x)^{2n} \over 64\pi n!\sqrt{nx+1+\xi ^2}}
D {du\over (1+u)^2}\eqno(7)
$$
\par
$$
D=
-2\eta ^2((1+\xi ^2)(u_+-u)(u-u_-))^n+{x^2n^2u^2\over 1+u}
((1+\xi ^2)(u_+-u)(u-u_-))^{n-1}
$$
  For $n=1$, integrating (7) by $u$, we get the cross-section of creation,
  calculated earlier in \cite{Pol1}:
$$
\sigma ={1\over 2}{\alpha g_{Xee}^2\over m_e^2}{1\over x}
\Biggl \{ \Biggl(1-2{\eta ^2\over x}+{2\eta^2(\eta^2-2)\over x^2}
\Biggr)\ln \Biggl({4(x+1)\over(2+x-\eta^2+\sqrt{(x-\eta^2)^2-4\eta^2}}\Biggr)\Biggr.
$$
$$
\Biggl.
+ \sqrt{(x-\eta^2)^2-4\eta^2}\Biggl( 1-{7\over 2}\eta^2 +{3\over 2}x
-8{\eta^2 \over x}-7{\eta^2\over x^2}\Biggr)\Biggr\}
$$
       At $x=5$  the cross-section is:
$$
\sigma \approx g_{aee}^2\cdot 5.4\cdot 10^{-24} cm^2
$$
        For numerical calculation, the physical parameters characterizing
the region of laser conversion are chosen in accordance with the projects
PLC  \cite{Gin, Tel2,Brink}.
        In the conversion scheme, it has been proposed to use a solid-state
laser with laser photon energy $\omega _{0}=1.17$ eV.
The length of conversion region characterized by a high density
of laser photons is  $l \sim 0.15$ á¬,which value is close to the length
of the electron bunch.
The invariant mass of the $\gamma _{0}e$--system for the energy of electron
$E_e=250$ GeV is comparable with the electron mass
$W_{\gamma _{0}e} \approx 1.21$ ŒeV ($x=4.5$).
It is convenient to express $\xi ^2$ in terms of the energy $A$,
the duration $\tau $ and radius $a_{\gamma }$ of the laser flash
in the interaction point
$$
\xi ^{2} = {A \over A_{*}},\quad {\rm where}\quad A_{*} =
{\tau c \over 4} \cdot \left ({ m_{e} \omega _{0} a_{\gamma} c \over
e \hbar } \right ) ^{2}.
$$
\par
\noindent
For the values $\pi a_{\gamma }^{2} \approx 10^{-5} \quad cm^2$,
$A_{*} = 100 $ J and at the energy of laser flash $A = 25$ J
${\xi}^{2} =0.25$.
Really the value ${\xi}^{2}$ can be reach 0.6.
The number of produced Goldstones is
$$
N_X={N_e\tau \over 2} \sum_{n>n_{th}} \int _{u_-}^{u_+}dW_n\eqno(8)
$$
  The Goldstone energies are distributed in the interval
$$
{\eta ^2\over (x+\eta ^2)R}E_e<\epsilon _X<E_e {x+\eta ^2\over x+1}R,
{\rm where}\, R={1\over 2}\left (1+\sqrt{1-{4\eta ^2(x+1)\over (x+\eta ^2
)^2}} \right )\eqno(9)
$$

For $E_e=250$ GeV at  $m_X=10$ KeV they correspond to the interval
 $$20\, MeV<\epsilon _X< 208\,GeV$$.
Since the effective mass of the $\gamma _0 e$--system not large
the characteristic emission angles of Goldstones relative to the
direction of the electron bunch are $\leq m_e/E_e \approx 10^{-5}$.
Therefore the angular spread of Goldstones is defined by the angular
spread of electrons in the beam  ($\approx 10^{-4}$).

   The numerical calculation of the number of Goldstone particles
and the couplings are shown in the table.
\vspace{0.5cm}

\centerline{\large Table}

\begin{center}
\begin{tabular}{|c|c|c|c|}\hline
 & $g_{Xee}$ & Cross-section $\sigma (cm^2)$ & The number of events per year\\
\hline  Standard axion & $2\cdot 10^{-6}$ &$2.2\cdot 10^{-35}$ &$7\cdot 10^{10}$ \\
\hline  "Invisible" axion &$3\cdot 10^{-8}$ &$4.9\cdot 10^{-39}$ &$1.5\cdot 10^7$ \\
\hline  Arion & $2\cdot 10^{-6}$ & $2\cdot 10^{-35}$ &$7\cdot 10^{10}$ \\
\hline  Familon & $1.4\cdot 10^{-13}$ &$1\cdot10^{-49}$&$3\cdot10^{-4}$\\
\hline  Majoron & $3.4\cdot 10^{-14}$ &$5\cdot 10^{-51}$ & $1.5\cdot 10^{-5}$\\
\hline \end{tabular} \end{center}
      The numerical estimations show that by obtaining intensity high
energy photons the large number of light Higgs particles will generate.

\section{The registration}

To registarate the new particles it is necessary to build the special
detectors.
For detection it is proposed to use production of hadrons in collisions of
 produced Goldstones with nuclei of the pin-type lead rod with radius
$\approx 2$ cm and length $\approx 100$ m , placed after a shield to get rid
of the background:
   $$
   X + Pb \to h \quad {\rm ( ¤à®­ë) } \eqno(10)
   $$
   \par
   This reaction will be observed as the production of hadron jets with
total energy $\sim \epsilon _X$ and transverse momentum
$p_\perp \sim 300$ MeV/c.
\footnote{The cross-section of lepton pairs production in Bete-Haitler
reaction $X+Pb \to e^+e^-+ \cdots ,X+Pb \to \mu ^+\mu ^- +\cdots $
approximately an order less then cross-section (10)}.
  Let us evaluate the number of events as an example of  standard axion.
        The cross-section of reaction (10) is $\sim N$ times as large
as the cross-section of the axion--nucleon interaction, $\sigma _{an}$,
where $N$ is the number of nucleons in a nucleus.
        The cross-section  $\sigma _{an} \approx
 f_{a\pi }\sigma _{\pi n}(v/f_{pq})^2$,where $f_{a\pi }$ the amplitude
of the axion--pion transition for the standard axion, $f_{a\pi }=
2\cdot 10^{-4}$\cite{SW}.
Therefore
$$
\sigma _{a+Pb \to h} \approx Nf_{a\pi}^2(v/f_{pq})^2\sigma _{\pi n}
\approx 5\cdot 10^{-34}cm^{2}.
$$
\par
\noindent
 It  corresponds to the fact that on the path of lead of 100 m will be
observed one event (2), (10) per hour.
 The increase of additional $U(1)$ symmetry scale on the order
 give the decrease of the events number in lead of 4 orders,
because we really think it is possible to reach $f_{pq} \sim 10$ TeV in this experiment.
        The background for the reaction (2), (10) will be produce via
   the high energy photon interaction with the matter of detector.
   These neutrinos are obtained from hadrons decay between collisions.
   The energy of creating of such away a neutrinos will be less than
   $\epsilon _X$ and their angle spread will be wide enough,
because the background  is weak in the pin-type rod.

\section{Acknowledgments}

       The author is grateful to I.F.Ginzburg for numerous discussion and
also to V.A.Naumov for benevolent remarks.

\end{document}